# 1-octadecene monolayers on Si(111) hydrogen-terminated surfaces: Effect of substrate doping

## Corinne MIRAMOND [1] and Dominique VUILLAUME [2]

**Institut d'Electronique, Micro-électronique et Nanotechnologie, CNRS, BP69, Avenue Poincaré**
**F-59652 cedex, Villeneuve d'Ascq, France**

## Abstract

We have studied the electronic properties, in relation to their structural properties, of monolayers of 1-octadecene attached on a hydrogen-terminated (111) silicon surface. The molecules are attached using the free-radical reaction between C=C and Si-H activated by an ultraviolet illumination. We have compared the structural and electrical properties of monolayers formed on silicon substrate of different types (n-type and p-type) and different doping concentrations from low-doped ($\sim 10^{14}$ cm$^{-3}$) to highly doped ($\sim 10^{19}$ cm$^{-3}$) silicon substrates. We show that the monolayers on n-, p- and p$^+$-silicon are densely packed and that they act as very good insulating films at a nanometer thickness with leakage currents as low as $\sim 10^{-7}$ A.cm$^{-2}$ and high quality capacitance-voltage characteristics. The monolayers formed on n$^+$-type silicon are more disordered and therefore exhibit larger leakage current densities ($> 10^{-4}$ A.cm$^{-2}$) when embedded in a silicon/monolayer/metal junction. The inferior structural and electronic properties obtained with n$^+$-type silicon pinpoint the important role of surface potential and of the position of the surface Fermi level during the chemisorption of the organic monolayers.

---

[1] Present address: ST Microelectronics, Crolles, France
[2] Corresponding author: vuillaume@isen.iemn.univ-lille1.fr

# I. INTRODUCTION

The functionalization of a semiconductor surface by the chemisorption of a monolayer of organic molecules is of great interest for a wide range of technological applications. These applications concern wetting, adhesion, lubrication, chemical and biological sensors and electronic devices. For the latter, organic monolayers are used to modify the electronic properties of metal and semiconductor surfaces,[1,2] to pattern these surfaces with a high resolution by electron-beam lithography[3] and they are also incorporated as parts of nanometer-scale electronic devices.[4,5] A standard technique in order to fabricate monolayers on silicon wafer uses the chemical reactivity of molecules bearing a trichlorosilane head group which are chemisorbed on a naturally oxidized, hydroxyl-rich, silicon surface.[6-9] The highly insulating monolayers of alkyltrichlorosilane molecules made by this process[10,11] can be incorporated as gate insulators in nanometer-scale field-effect transistors (gate length 25-30 nm) with both inorganic (silicon) and organic semiconductors.[4,5] However, the presence of the ultra-thin native oxide (typically 1-1.5 nm thick) precludes the study of a true silicon/organic interface, the structural and electrical properties of this native oxide being not well controlled (high density of electrical active defects). Recently, Linford and Chidsey demonstrated that high quality, densely-packed, monolayers can be prepared from 1-alkenes on hydrogen-terminated silicon(111) surfaces.[12,13] Direct evidence of Si-C bonding was shown using X-ray photoelectron (XPS) spectroscopy and photoelectron diffraction.[14,15] Following this pioneering work, several routes were described to attach molecules (mainly long alkyl chains) on hydrogen-terminated single-crystal silicon (111) and (100) surfaces and porous silicon substrates. This includes: thermally activated chemical reactions,[13,16-20] reaction with organomagnesium or organolithium,[14,15,21-25] photochemically (Ultra Violet-light) activated process,[12-14,26,27] hydrosilylation[21,28,29] and electrochemical process.[30-33]

The alkene monolayers are thermally stable in vacuum up to about 615 K.[18] Moreover, several works have shown that these alkene monolayers act as a good surface passivation with a low interface state density ($\sim 10^9$ - $10^{11}$ cm$^{-2}$eV$^{-1}$),[34,35] a long minority carrier lifetime ($\sim$130 to 300$\mu$s)[34,36] and a low surface recombination velocity (<200 cm.s$^{-1}$ and down to $\sim$25 cm.s$^{-1}$).[16,34,36,37] All these features indicate that these monolayers are suitable candidates for silicon passivation layers and that they should eventually be used as ultra-thin gate dielectrics in nanoelectronics devices. In all of these studies, little attention was paid to the type of substrate doping (n-type vs. p-type doped silicon) and the doping concentration (intrinsic semiconductor, i.e. with a low doping concentration of about $10^{10}$ cm$^{-3}$, vs. highly doped, also called degenerated, semiconductor, i.e. about $10^{19}$-$10^{20}$ cm$^{-3}$). In order to foresee possible applications in nanoelectronics devices, the doping type and concentration of the substrate play a key role. An equally well passivation is required whatever the nature and level of the silicon doping.

In this work, we have studied the electronic and the structural properties of monolayers of 1-octadecene attached on hydrogen-terminated (111) surface of silicon by the free-radical reaction between C=C and Si-H, activated by ultraviolet illumination. We have compared the structural and electrical properties of monolayers formed on silicon substrate of different types (n-type and p-type) and different doping concentrations from low-doped ($\sim 10^{14}$ cm$^{-3}$) to highly doped ($\sim 10^{19}$ cm$^{-3}$) silicon substrates. We show that on n-, p- and p+-silicon, they are densely packed and that they act as good insulating films at a nanometer thickness with leakage currents as low as $\sim 10^{-7}$ A.cm$^{-2}$ when sandwiched in a silicon/monolayer/metal junction. They also have high quality capacitance-voltage characteristics. However, monolayers on n$^+$-type silicon are less packed, with a resulting leakage current >$10^{-4}$ A.cm$^{-2}$. The results are related to the important role of surface potential and the position of the surface Fermi level during the chemisorption of the organic monolayers.

## II. EXPERIMENTAL SECTION

### A. Sample preparation.

We used n-type (doping ~$3\times10^{16}$ cm$^{-3}$, resistivity $\rho$~ 0.23-0.27 $\Omega$.cm), n$^+$-type (~$10^{19}$ cm$^{-3}$, $5\times10^{-3}$ $\Omega$.cm), p-type (~$6\times10^{14}$ cm$^{-3}$, ~20 $\Omega$.cm and ~$2\times10^{16}$ cm$^{-3}$, 1-2 $\Omega$.cm) and p$^+$-type (~$10^{19}$ cm$^{-3}$, $10^{-2}$ $\Omega$.cm) silicon substrates. The hydrogen-terminated Si(111) surface was prepared as follows in a microelectronics grade clean room and using microelectronics grade chemicals. The samples were degreased ultrasonically in acetone and then in isopropyl alcohol. The native oxide was stripped by etching (30 sec) in 5% aqueous HF, then the surface was slightly re-oxidized (sacrificial oxide) by a H$_2$SO$_4$/H$_2$O$_2$ (3/1) treatment at ~100°C for 10 min (these two steps were repeated twice). *Caution: piranha solution (H$_2$SO$_4$/H$_2$O$_2$) is very exothermic and reactive with organics; it should be handled with extreme care*. This oxide was again stripped by HF etching and the Si surface was hydrogenated by treatment in 40% aqueous NH$_4$F (at pH 8) for 3 min. Rinsing in deionized water (18 M$\Omega$) followed each step. Alkyl monolayers were then formed by UV illumination from 1-octadecene on hydrogen-terminated Si(111) surface following the process described by Linford and Chidsey.[12,13] 1-octadecene (C$_{18}$H$_{36}$, 99.8% from Sigma-Aldrich) was placed into a glass vial and deoxygenated under reduced pressure ($5\times10^{-5}$ mbar) and constant stirring for 4 hours, the system is connected to a liquid-nitrogen-trapped diffusion pump. A few minutes before the end of this deoxygenation, a freshly prepared hydrogen-terminated silicon wafer was placed into a quartz vial isolated elsewhere in the vacuum line, and the quartz tube was evacuated to a residual pressure of $5\times10^{-5}$ mbar. The tube and the glass line near the tube were cooled in a dry ice/acetone bath. Following this, the whole system was isolated from the vacuum pump. The vial containing the 1-octadecene was heated in an oil bath, the 1-octadecene was distilled and purified through a short path distillation apparatus and vacuum transferred onto the quartz tube to completely wet the silicon sample. The sample is then illuminated by an ozone-free ultra-violet (UV) lamp ($\lambda$ = 253.7 nm, power 12 mW.cm$^{-2}$) placed at ~1cm from the quartz tube. A flow of fresh air is provided around the quartz tube to prevent heating. After 2 hours of illumination under reduced pressure, alkyl-terminated silicon surfaces were obtained. The quartz tube was opened to air, the 1-octadecene is decanted and the sample was removed. The sample was rinsed in a sonicated $CH_2Cl_2$ bath and stored in a dessicator until measurements.

For the electrical measurements, we formed the silicon/monolayer/metal (SMM) heterostructures by evaporating metal (aluminum and gold) through a shadow mask (electrode area: $10^4$ µm$^2$). To avoid any contamination of the surface during the metallization, an ultra-high vacuum (UHV) e-beam evaporation system was used. It was checked that a $10^{-8}$ torr vacuum is innocuous to the monolayers. To minimize the damage which may occur during this metallization, we used a low evaporation rate (0.1 to 0.5 Å/s), a large distance between the sample and the crucible of metal (about 70 cm), the sample temperature was maintained at 20°C and we only used metals with a relatively low melting temperature (aluminum and gold).[38] The electrode thickness was in the range 200-300 nm.

**B. Measurement techniques**

*1. Contact angle and wetting.*

The quality of the monolayers was first analyzed by measuring the water contact angle and the surface tension $\gamma$ of their surfaces. We used the extrapolation technique of Zisman [39] in which contact angles for sessile drops of homologous apolar liquids are plotted as a function of their liquid-vapor interfacial energy $\gamma_{LV}$. This technique gives the critical surface tension $\gamma_c$. The accuracy is estimated to be ± 0.5 mN/m. Water contact angles are measured using deionized water. We used a remote-computer controlled goniometer system (DIGIDROP by GBX, Lyon, France) for measuring the contact angles. The accuracy is ± 2°. All measurements were made in ambient atmosphere and at room temperature.

### 2. Infrared spectroscopy.

The hydrogen-terminated surfaces were analyzed, prior to monolayer formation, by Fourier transform infrared spectroscopy (FTIR) in transmission at Brewster angle on low-doped substrates, i.e. using a p-polarized beam incident on the surface at 73.7° (for silicon substrate) to the normal of the surface.[40] Measurements were performed in the external reflection mode (for highly doped $n^+$-type substrate) as described elsewhere.[41,42] After monolayer formation, we systematically measured the peak position of the symmetric and antisymmetric stretching modes of the $CH_2$ group. This is a useful tool to ascribe the degree of molecular order in the monolayers.[43] All trans extended alkyl chains have their $CH_2$ symmetric mode $\nu_s(CH_2)$ reported at 2846-2850 $cm^{-1}$ and the asymmetric mode $\nu_a(CH_2)$ at 2915-2920 $cm^{-1}$,[44-46] while conformationally disordered chains show blue shift of the two $CH_2$ stretching modes to ~2856 and 2928 $cm^{-1}$, respectively.[47] We used a SYSTEM2000 (Perkin-Elmer) FTIR spectrometer, equipped with a DTGS and a liquid nitrogen cooled MCT detectors. For all the spectra, 200 scans acquired at 2 $cm^{-1}$ resolution with a strong apodization, were co-added.

### 3. Atomic Force Microscopy and Conducting AFM.

We used Atomic Force Microscopy (AFM) to image the surface topography of bare Si-H surfaces and of the monolayer surfaces. A Nanoscope III (Digital Instruments) system in the tapping mode (TM-AFM) was used in air at room temperature. All images (512 x 512 pixels) were taken at the scanning rate of 2 - 2.44 Hz. Surface regions from 50 x 50 nm to 5 x 5 μm were imaged. We used a Conducting-AFM (C-AFM) to locally measure the current-voltage curve through the monolayer. We used an in-house modified Nanoscope III with a gold-coated $Si_3N_4$ tip. The contact force was controlled by the feedback loop of the Nanoscope, while the current-voltage curve was recorded using an external circuit. The current-voltage curve was acquired with an Agilent semiconductor parameter analyzer HP4145B. The current was first amplified by an in-house transconductance amplifier located nearby the tip and again amplified and filtered by a low-noise voltage preamplifier (Standford Research SR560) before to be recorded by the HP4145B. The detection limit is $10^{-14}$ A. The loading force for the C-AFM measurements was 10 nN.

### 4. Ellipsometry.

The monolayer thickness was measured by ellipsometry at 633 nm. In the calculation of the thickness, we used an isotropic value of $n_{mol}$=1.50 for the monolayer refractive index at 633 nm and 3.865 for the silicon substrate, values in the literature are in the range 1.45-1.50.[43] One can notice that a change from 1.50 to 1.45 results in less than 1 Å error. Accuracy of the monolayer thickness measurements is estimated to be ± 0.2 nm.

### 5. Electrical measurements.

The SMM structures were mounted onto a wafer chuck and contacted by precision micromanipulators. Electrical transport through the monolayers was determined by measuring the current density versus the applied dc voltage (-1 V to +1 V) with a HP4140B picoampmeter. We used a slow speed (10 mV/sec) step-like voltage ramp (step voltage 10 mV) in order to avoid any transient effects due to displacement current since the structure mainly behaves as a capacitor. Dynamic capacitance versus dc bias was measured from 100 Hz to 1 MHz (ac signal ~ 25 mV$_{eff}$) by HP4192A and HP4274A impedance meters. In both cases, dc and ac biases were applied on the metal counter-electrode, the silicon substrate being grounded. Measurements at room temperatures were done in a shielded dark box and in the ambient atmosphere.

## III. RESULTS AND DISCUSSION

### A. Hydrogen-terminated surfaces.

The hydrogen-terminated surfaces were analyzed by FTIR and TM-AFM before the formation of the monolayers. Characteristic peaks of Si-H vibrations are observed at 2084 cm$^{-1}$. We do not detect any Si-H$_2$, Si-H$_3$ peaks while the Si-0-Si stretching peak at 1060 cm$^{-1}$ has disappeared. TM-AFM images (Figure 1) showed the characteristic features of Si(111)-H surfaces (atomically flat terraces with step height of ~0.3-0.6 nm). We have not observed any difference in the topography of the Si-H surfaces whatever the nature and value of the doping. Figures 1 (a and b) show typical TM-AFM images for a n-type and n$^+$-type substrates. Water contact angles are 79±2° in all cases.

### B. 1-octadecene monolayers on n-, p- and p$^+$-type surfaces.

FTIR (p-polarized beam in transmission at Brewster angle, on low-doped substrates only) spectra showed the disappearance of the Si-H vibration peaks and the appearance (Fig. 2) of the characteristic peaks of methylene (CH$_2$) stretching modes at 2919 cm$^{-1}$ (asymmetric) and 2851 cm$^{-1}$ (symmetric) and of the methyl group (CH$_3$) at 2965 cm$^{-1}$ (asymmetric, ip) and 2880 cm$^{-1}$ (symmetric). We do not observe a significant re-appearance of the Si-O-Si peaks after the formation of the monolayer, but the spectra in this region are noisy and we cannot rule out a slight re-oxidation of the silicon (see XPS data below). The monolayer thickness (ellipsometry) is between 1.98 to 2.35 nm (table I) for all the monolayers (taking a refractive index of 3.865 for silicon and 1.5 for the alkyl monolayer) in good agreement with an array of densely packed alkyl chains in their all-trans configuration and tilted by $\theta \sim 30°$ from the surface normal[48] (we used the thickness formula $d = 0.126 \times (n-1) \times \cos\theta + 0.186 \sim 2.1$ nm for octadecene, n=18 carbon atoms, assuming that the C-Si and C-C bond length projected along the molecular axis are 0.184 and 0.126 nm, respectively[14]). The contact angle measurements give 101-105° for water, 40-44° for hexadecane and a critical surface tension (Zisman method for a series of linear alkanes) of 20.2-20.6 mN/m. These values suggest a monolayer surface made of a dense array of methyl groups. TM-AFM images showed that the topography of the monolayers (Fig. 1-c) reproduce the features of the underlying Si(111) surface. Finally, X-ray photoelectron spectroscopy (XPS) on freshly prepared monolayers (Fig. 3) showed the presence of C(1s), Si(2p) with a small amount of O(1s), $[O_{1s}]/[C_{1s}] \sim 0.04$ (corresponding to a residual oxygen concentration of $3 \times 10^{13}$ atoms/cm$^2$), but no discernible oxidized Si signal. All these results are in good agreement with previous reports [12-15] and indicate the formation of a densely packed 1-octadecene monolayer directly on silicon without the presence of an underlying oxide. After nine weeks stored at room temperature in air, we have observed a slight re-oxidation of the Si surface. The $[O_{1s}]/[C_{1s}]$ increases to 0.14, while the oxidized silicon peak at ~103 eV is observed. This ratio would correspond to about $1.3 \times 10^{15}$ oxygen atoms/cm$^2$, i.e. about one oxygen monolayer. Even though the silicon is slightly re-oxidized, it is probably not a continuous film, but some oxidized spots may be formed in defective regions of the monolayer.

## C. 1-octadecene monolayers on n[+]-type surfaces.

Hexadecane contact angle and surface tension are 29-40° and 21.7-23.1 mN/m, respectively, for monolayers on n[+]-type silicon, suggesting a less densely packed, more disordered, monolayer in that case. This lower values (compared to the case of a densely-packed, mainly $CH_3$-terminated monolayer) correspond to a surface made of a mixture of $CH_3$ and $CH_2$ as expected for a disordered monolayer.[43] The water contact angles are not changed, still in the range 101-104°. This is explained by the fact that water contact angles are less sensitive (at least to a certain extend) to surface disorder than hexadecane contact angles.[49] FTIR (p-polarized light, external reflection) spectrum shows "positive" and "negative" (Fig. 4) peaks as expected for such an external reflection infrared experiment on a silicon substrate. As clearly stated by Hoffman et al.[41,42] a highly ordered alkyl monolayer should exhibit strongly positive peaks at ~2920 and ~2850 cm[-1] (methylene asymmetric $\nu_a(CH_2)$ and symmetric modes $\nu_s(CH_2)$) and negative peaks (of much lower amplitude) at ~2968 and ~2879 cm[-1] arising from methyl vibration modes ($\nu_a(CH_3)_{ip}$ and $\nu_s(CH_3)$, respectively). A strongly disordered, liquid-like, film shows only negative peaks. In our case, the spectrum showed both positive peaks (at ~2921 and ~2857 cm[-1]) and several negative peaks (at ~2869, ~2935 and ~2971 cm[-1]) of almost the same amplitudes. Our spectrum strongly resembles that observed by Hoffman et al. (see Fig. 7-B in Refs. [42]) and also by Jeon et al. (see Fig. 5-B in Ref. [50]) for an intermediate case, where the monolayer contain an important concentration of conformationally disordered alkyl chains. Thus, from both FTIR and wettability we concluded that the 1-octadecene monolayers on a highly doped n[+]-type substrate were more disordered than those on the n-, p- and p[+]-type silicon substrates. Moreover, TM-AFM (Fig 1-d) showed that featureless images (lost of clear images of step edges) were also obtained for the monolayers on n$^+$-type silicon. This feature can be ascribed as due to the presence of a disordered monolayer which hidden the underlying topography of the silicon surface.

## D. Electrical properties.

When a macroscopic metal electrode is evaporated on top of an organic monolayer, it is well known that the amplitude of the leakage current through the monolayer is dependent on the number of metallic filamentary pathways present in this monolayer underneath the electrode.[51,52] These metallic filamentary pathways results from metal diffusion through the monolayer, metal-related defects created by the impacting metal atoms, or by the filling with metal of pre-existing defects in the monolayers (e.g. pinholes). The more disordered the monolayer the greater the number of metallic filamentary pathways and hence the higher the measured current. The measure of the leakage current is a very sensitive tool to estimate the degree of disorder in the monolayer. For example, we have shown elsewhere,[11] on a deliberately disordered self-assembled alkylsiloxane monolayer, that less than few % of disorder in the monolayer induces an increase in the current by a factor of about $10^4$. Since these defects are located at random in the monolayers, the number of defects under the electrode may vary from sample to sample, and a statistical analysis of the leakage current has to be carried out. Not only the minimum leakage current depends on the quality of the monolayer, but also the average value and the distribution.

Figure 5 shows the best-case (i.e. lowest) current density-voltage (J-V) characteristics for the SMM junction formed on the 4 different types of substrate. The results from the control sample (Al electrode directly evaporated on freshly prepared Si(111)-H surface) are also presented as a comparison (J saturates at $10^2$ A.cm$^{-2}$ due to current compliance of the equipments). The currents are strongly decreased (compared to the control samples) for the more densely packed monolayers (those on n-, p- and p$^+$-type silicon substrates). Current densities as low as $10^{-7}$-$10^{-6}$ A.cm$^{-2}$ were obtained at 1 V, on a par with those previously demonstrated for monolayers of alkyltrichlorosilanes (on naturally oxidized n$^+$-type silicon).[10,11] The lowest current achieved for the monolayer on n$^+$-type silicon is only ~$10^{-4}$ A.cm$^{-2}$ (at 1V). The current density histograms for monolayers on the 4 types of substrates are shown in figure 6. It is clear that the distribution for the monolayers on n$^+$-type silicon is centered at higher current densities. For the three better cases (n, p and p$^+$), the width of these distributions are also comparable with that observed for alkyltrichlorosilane monolayers on naturally oxidized silicon surfaces.[4]

To eliminate the role of defects in the conduction mechanism through the monolayer and to measure the electronic properties at the nanometer-scale, we used C-AFM.[53] For the monolayers on n-, p- and p$^+$-type silicon substrates, a typical current density between $10^{-8}$ and $10^{-4}$ A.cm$^{-2}$ (see Fig. 6) would correspond to a C-AFM current always < $10^{-16}$ A assuming a contact area of 10 to $10^2$ nm$^2$ for our C-AFM tip. Such a current is below the detection limit of our C-AFM ($10^{-14}$ A). Indeed, we did not detect any current (fig. 7) on these samples by C-AFM (for bias in the range -1V to +1V). This is in agreement with previously reported C-AFM measurements on alkylthiol monolayers on gold.[53,54] If we extrapolate such data (taken for short chains $\leq$ 12 CH$_2$ groups) to longer chain with 17 CH$_2$, the resistance is of the order of $10^{15}$ $\Omega$ (i.e. $10^{-15}$ A at ±1V). Of course, this agreement concerns only the order of magnitude since they might be some differences in the tip size and loading force (current in C-AFM depends on the loading force) between these different experiments. Moreover, we compare the conduction through alkyl chains attached to gold via a sulfur atom (S-Au bond) and attached to silicon via a Si-C bond. It is also clear that the electrical conduction in an electrode/molecules/electrode junction depends on the nature of the chemical link between the molecule and the electrode,[55-57] and thus such a comparison is only qualitative. It establishes that our 1-octadecene monolayers on silicon have a similar insulating behavior than others alkyl chains on other substrates. We only observed a sudden increase in the current at bias higher than 2 V, probably corresponding to dielectric breakdown.[54,58,59] In this n-type silicon substrate, the voltage drop in the silicon for a positive voltage applied on the metal gate is weak (~170 meV corresponding to the difference between the bottom of the conduction band and the Fermi energy for such a n-type silicon in accumulation regime) and thus the estimated breakdown field is about 8.7 MV/cm for these monolayers (taking a monolayer thickness of ~2.3 nm, see Table 1). This value is close (a factor 2.3) to other reports (20 MV/cm) for alkylthiol on gold[54] and alkene on silicon.[59] Considering the case of monolayers on a $n^+$-type silicon, a current density of $10^{-1}$ A.cm$^{-2}$ (as shown in Fig. 6) would correspond to a current in the range $10^{-14}$-$10^{-13}$ A. If the high current density measured for the monolayers on the $n^+$-silicon at a macroscopic scale is intrinsic to the monolayer structure and not due to metallic filamentary pathways formed during the electrode evaporation, this current should be detected by the C-AFM measurements. No measurable current (in the voltage range -1V to +1V) was detected by C-AFM for our monolayers on the $n^+$-silicon substrate (Fig. 7). This confirms that the monolayers on the $n^+$-silicon are less densely packed, as revealed by IR and contact angle measurements, and consequently that they are more sensitive to the metal-related defects formation during the top electrode evaporation. Thus the higher current measured through these monolayers on the $n^+$-type silicon are mainly due to a larger number of filamentary metallic pathways, while the intrinsic conductivity of the molecules at the nano-scale are almost the same as for monolayers on the other types of silicon substrates.

Figure 8 shows capacitance-voltage (C-V) curves measured at 1 MHz for monolayers on n- and p-type silicon substrates (measured with evaporated electrodes). Carrier accumulation, depletion and inversion regimes are clearly obtained for these SMM capacitors.[60] More precisely, the formation of a strong inversion layer at the silicon/organic monolayer interface is clearly established by the C-V measurements under white light illumination. These good capacitance results are the fingerprint of the strong insulating properties of the organic monolayer and low silicon surface state density. The values of capacitance in accumulation were found to be near the expected values (taking an averaged monolayer thickness of 2.3 nm and a dielectric constant between 2.2 and 2.5). From the C-V characteristics at high and low-frequencies and from detailed admittance spectroscopy, we have shown elsewhere that the state density ($D_{it}$) at the (111)Si/1-octadecene interface in our samples is about 2-3x10$^{11}$ cm$^{-2}$eV$^{-1}$ at mid-gap.[35] We notice that this value is larger than the trap concentration ($N_{it}$)[61] of about 3x10$^9$ cm$^{-2}$ reported elsewhere for alkylated silicon surfaces.[34] This can be due to a slight re-oxidation of the surface (see our XPS data, figure 3) during the time elapsed in air between the sample preparation and the electrical measurements ($D_{it}$ for a native oxide is 10$^{12}$-10$^{13}$ cm$^{-2}$eV$^{-1}$). It is clear from the comparison to trap densities measured elsewhere[34] that a part of the ~10$^{11}$ trap/cm$^2$ measured in our sample comes Si/SiO$_2$ interface defects re-introduced during the slight re-oxidation. Nevertheless, such a value represents a low defect density for a Si(111) interface compared to typical densities of about 10$^{13}$ cm$^{-2}$eV$^{-1}$ commonly observed for the Si(111)/SiO$_2$ interface (not submitted to any high temperature post-oxidation anneal as usual in silicon technology).[62-64] The nearest-neighbour distance between Si atoms on the (111) surface is 3.84Å, a distance smaller than the diameter of the alkyl chain (4.8 Å) and thus the Si atoms are not all passivated by alkyl chains. About 50% of the surface silicon atoms (a silicon surface contains ~10$^{15}$ atoms/cm$^2$) are passivated by the 1-alkene molecules.[48] The rest (to reach the residual defect density of ~$10^{11}$ cm$^{-2}$) is passivated by H (or OH after the slight re-oxidation).

**E. Discussion and conclusion**.

A striking feature of the experimental data is the poor structural and electrical quality systematically observed for monolayers formed on n$^+$-type silicon. This may be explained as follows. According to the proposed mechanism,[13] a radical site (the silicon dangling bond) is formed under UV irradiation and then react with the C=C double bond to form a surface-bonded alkyl radical (through a Si-C bond as evidenced recently[14,15]). This radical may abstract a hydrogen atom from a neighboring Si-H to saturate the alkyl chain, leaving a new dangling bond at the surface and so on. Thus, this reaction requires unsaturated Si dangling bonds at the surface. The Si(111) dangling bond is well known as an amphoteric defect with three possible charge states (positive with no electron on the dangling orbital, neutral when occupied by one electron, and negative when doubly occupied by two electrons). The respective ionization energy levels in the Si band gap are well known: at ~0.3 eV above the valence band edge for the +/0 transition and at ~0.3 eV below the conduction band edge for the 0/- transition.[62,63] On a highly doped, degenerate n$^+$-type substrate, the surface Fermi level lies very close to the conduction band, and almost all of the Si dangling bonds may be saturated with 2 electrons, thus rendering the formation of a Si-C bond by addition of the 1-octadecene on this Si-based radical more difficult than on p$^+$-, p- and n-type silicon substrates for which the Fermi level at the Si surface is not so close and the dangling bonds are mainly single electron occupied. As the Si-based radicals become more nucleophilic (due to the n$^+$ doping), the rate of addition to 1-octadecene decreases. In the same time, this Si-based radical should react faster with electrophilic impurities possibly present in the neat 1-octadecene.

A larger incorporation of impurities in the monolayer would also result in a more disordered one.

In conclusion, we have shown that 1-octadecene monolayers formed on hydrogen-terminated Si(111) (of n-, p- and p$^+$-type) by the free-radical reaction between C=C and Si-H activated by UV illumination are densely packed and can act as very good insulating films at nanometer thicknesses. Even if a large dispersion of the leakage currents has been observed through the silicon/monolayer/metal junction (which requires a further improvement of the monolayer quality), the leakage currents are within the range $10^{-7}$-$10^{-4}$ A.cm$^{-2}$. However, more disordered monolayers having higher leakage currents are systematically obtained for the monolayers formed on highly doped n$^+$-type silicon. Due to energy position of the surface Fermi level during the chemisorption on this n$^+$-type silicon, this feature can be presumably ascribed to the anionic character of the silicon dangling bond on this n$^+$ surface, and consequently to a more difficult reaction of the 1-octadecene.

**Acknowledgements.** We would like to thank X. Wallart for the XPS measurements. We thank also R. Boukherroub and S. Arscott for the critical reading of the manuscript. A part of this work was supported by the IFCPAR (Indo-French Center for the Promotion of Advanced Research) under contract n°1614-1.

Table I : Water and hexadecane contact angles, Zisman critical surface tension and monolayer thickness (ellipsometry) for the different samples made on silicon with a large range of doping. These numbers are averaged values. Typically, we measured contact angles and thicknesses at five different places on the sample surface ("n.m." means "not measured").

| Silicon substrate and doping (cm$^{-3}$) | Sample # | Water contact angle (°) | Hexadecane contact angle (°) | Surface tension (mN/m) | Monolayer thickness (Å) |
|---|---|---|---|---|---|
| N$^+$, $10^{19}$ | #lot6 | 101.6 | 31.0 | 21.8 | 22.7 |
| N$^+$, $10^{19}$ | #lot7 | 102 | 39.4 | 22.8 | 19.9 |
| N$^+$, $10^{19}$ | #lot9 | 102 | 28.7 | 23.1 | 21.1 |
| N$^+$, $10^{19}$ | #lot11b | 104 | 35.8 | 21.7 | 21.6 |
| N, $3 \times 10^{16}$ | #lot13 | 105 | 44.0 | 20.2 | 21.7 |
| N, $3 \times 10^{16}$ | #lot16 | 103.5 | 42.2 | 20.6 | 22.0 |
| P, $6 \times 10^{14}$ | #lot5 | 101 | 40.3 | 20.3 | 20.1 |
| P, $6 \times 10^{14}$ | #lot10 | 105 | 42.6 | 20.3 | 19.8 |
| P, $6 \times 10^{14}$ | #lot11a | 102 | 40.8 | 20.4 | 23.5 |
| P, $2 \times 10^{16}$ | #lot12 | 105 | 44 | 20.2 | 22.0 |
| P$^+$, $10^{19}$ | #lot19 | n.m. | 41.7 | 20.2 | n.m. |

# FIGURE CAPTIONS

Figure 1 : TM-AFM images (all 2µm x 2µm). (a) Bare Si-H surface, n-type silicon, (b) Bare Si-H surface, $n^+$-type silicon, (c) 1-octadecene monolayer on n-type silicon, (d) 1-octadecene monolayer on $n^+$-type silicon.

Figure 2 : FTIR (Brewster angle in transmission, p-polarized) spectrum of a 1-octadecene monolayer on a low-doped ($6 \times 10^{14}$ $cm^{-3}$) p-type Si substrate.

Figure 3 : XPS spectra of a fresh 1-octadecene monolayer and after a period of nine weeks stored in ambient air.

Figure 4 : FTIR (external reflection, p-polarized) spectrum of a 1-octadecene monolayer on a $n^+$-type substrate.

Figure 5 : Leakage current density versus voltage characteristic (best-case) though the monolayers formed on various silicon substrates (n-, p-, $n^+$- and $p^+$-type). The substrate is grounded and the voltage is applied on the metal electrode. The saturation at $10^{-2}$ $A.cm^{-2}$ for the control sample (no monolayer) is due to the current compliance of the picoampmeter (Agilent HP4140B).

Figure 6 : Statistical distribution of the leakage current measured at ±1V for the monolayers formed on different silicon substrates. For the highly doped substrate ($n^+$), the double distribution (peaks at ~$10^{-3}$-$10^{-4}$ $A.cm^{-2}$ and at ~$10^{-1}$ $A.cm^{-2}$) corresponds to different series of samples with two different surface tensions, samples #lot6 and #lot11b (with $\gamma$~21.7 mN/m) and samples #lot7 and #lot9 (with $\gamma$~23 mN/m), respectively. This feature also illustrates the relationship between increased disorder (as indicated by the higher $\gamma$ value) and increased leakage current.

Figure 7 : Current-voltage measured by C-AFM on a 1-octadecene monolayer on a n-type silicon substrate. The loading force is 10 nN. The inset shows that only noise current is measured between 0 and 1V.

Figure 8 : (a) High frequency (1 MHz) capacitance-voltage characteristics for monolayers formed on n- (■) and p-type (●) silicon substrates showing the typical accumulation (acc), depletion (dep) and inversion (inv) regimes. The substrate is grounded and voltage is applied on the metal counter electrode. The dotted lines represent the capacitance in accumulation regime ($C_{acc}=\varepsilon.\varepsilon_0/d$) calculated with a monolayer thickness d = 2.3 nm and a relative dielectric constant $\varepsilon$ = 2.2 and 2.5 ($\varepsilon_0$ is the vacuum dielectric constant). For monolayers on $n^+$- and $p^+$-type substrate, the capacitance-voltage characteristics are rather flat at the $C_{acc}$ value due to the highly-degenerated character of the silicon. (b) High frequency (1 MHz) capacitance-voltage characteristics in dark (●) and under white light illumination (■) for a monolayer formed on a p-type silicon substrate. The large increase of the capacitance in inversion is the fingerprint of the light-induced formation of the inversion layers (populated by electron from light-induced electron-hole pair generation) at the silicon/octadecene interface.

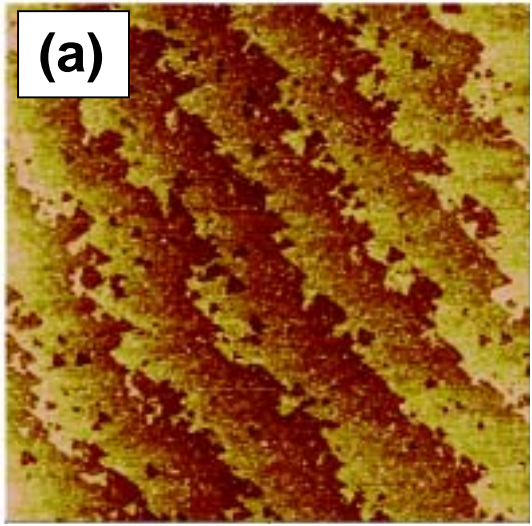
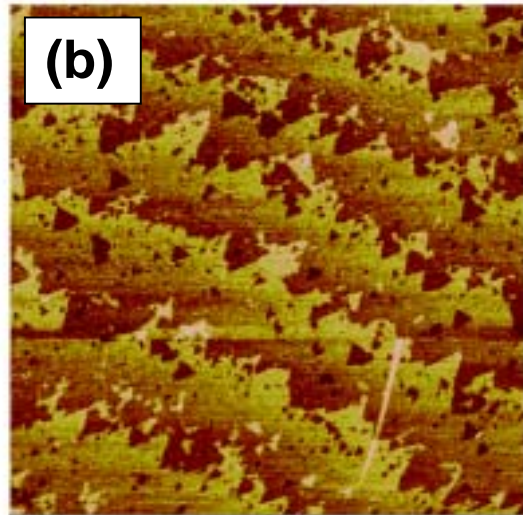
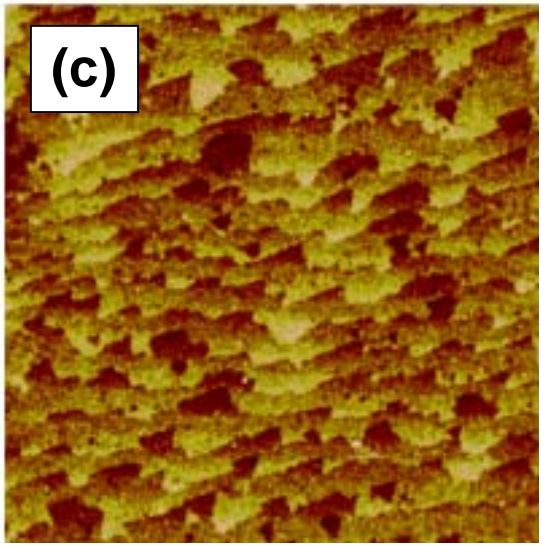
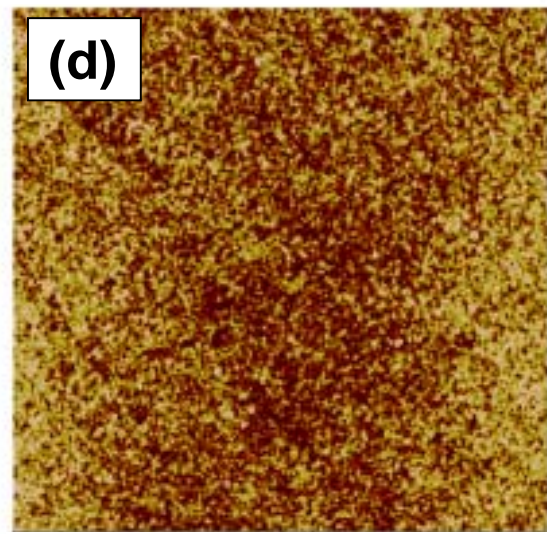

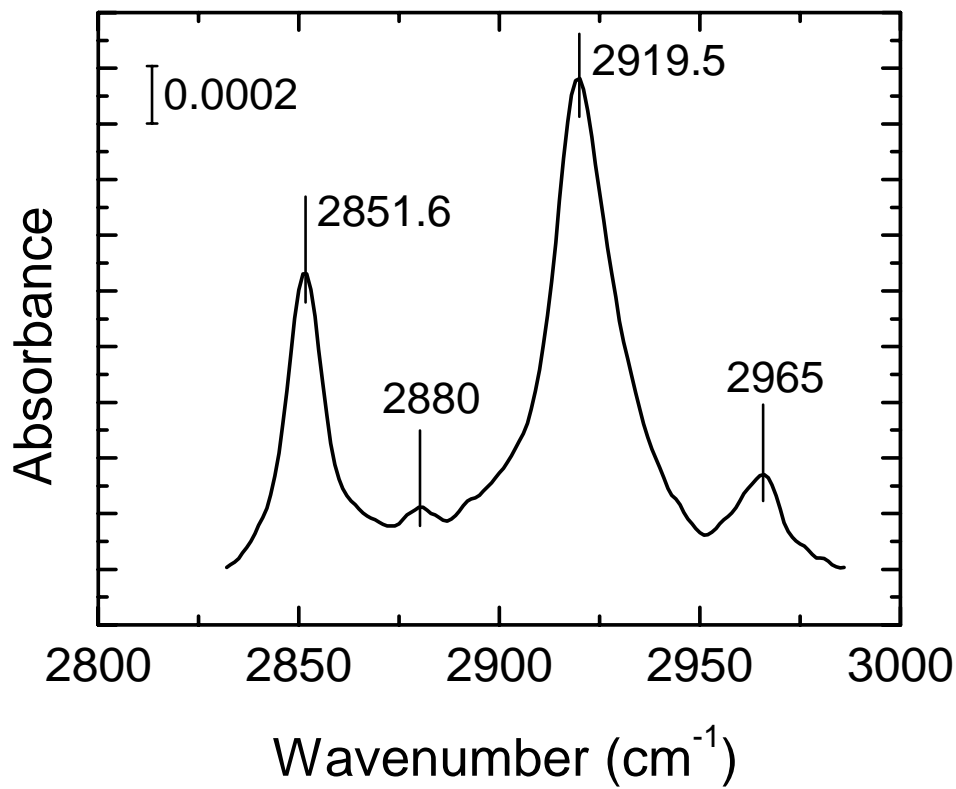

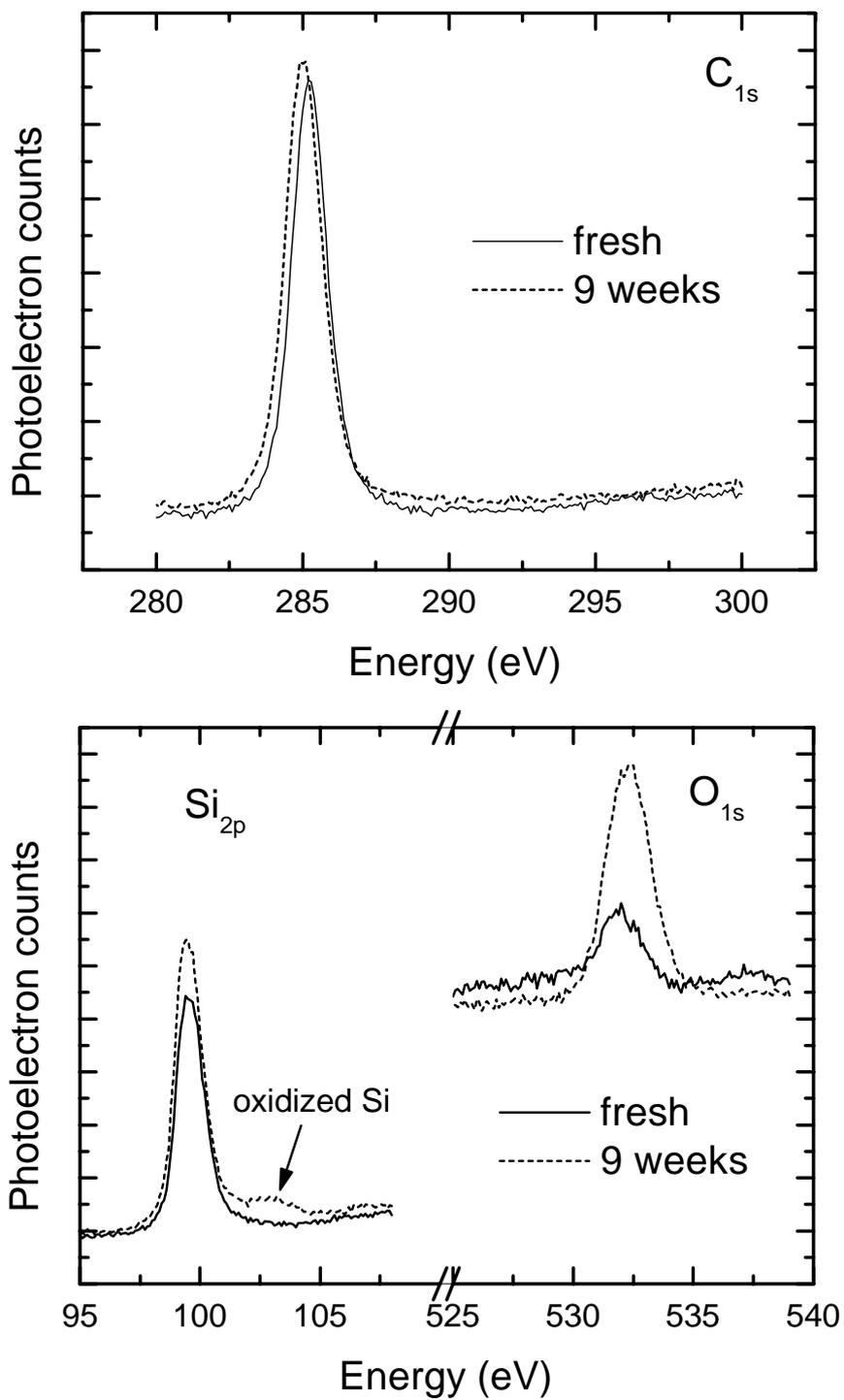

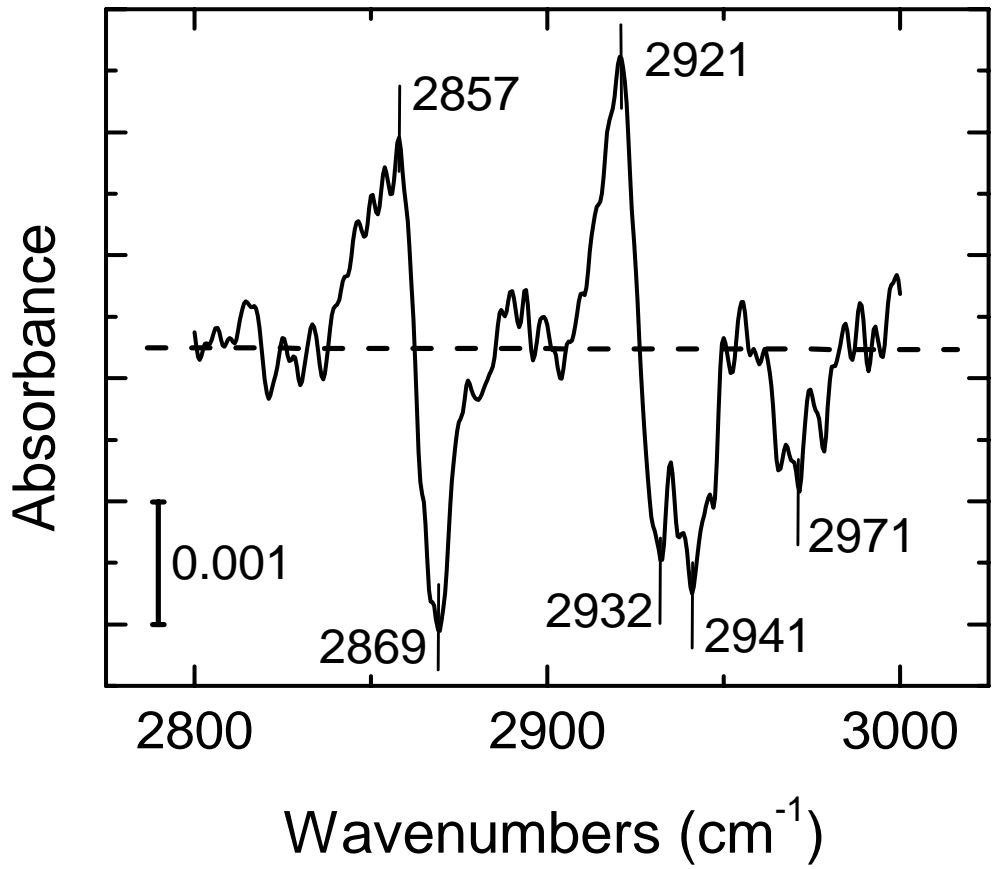

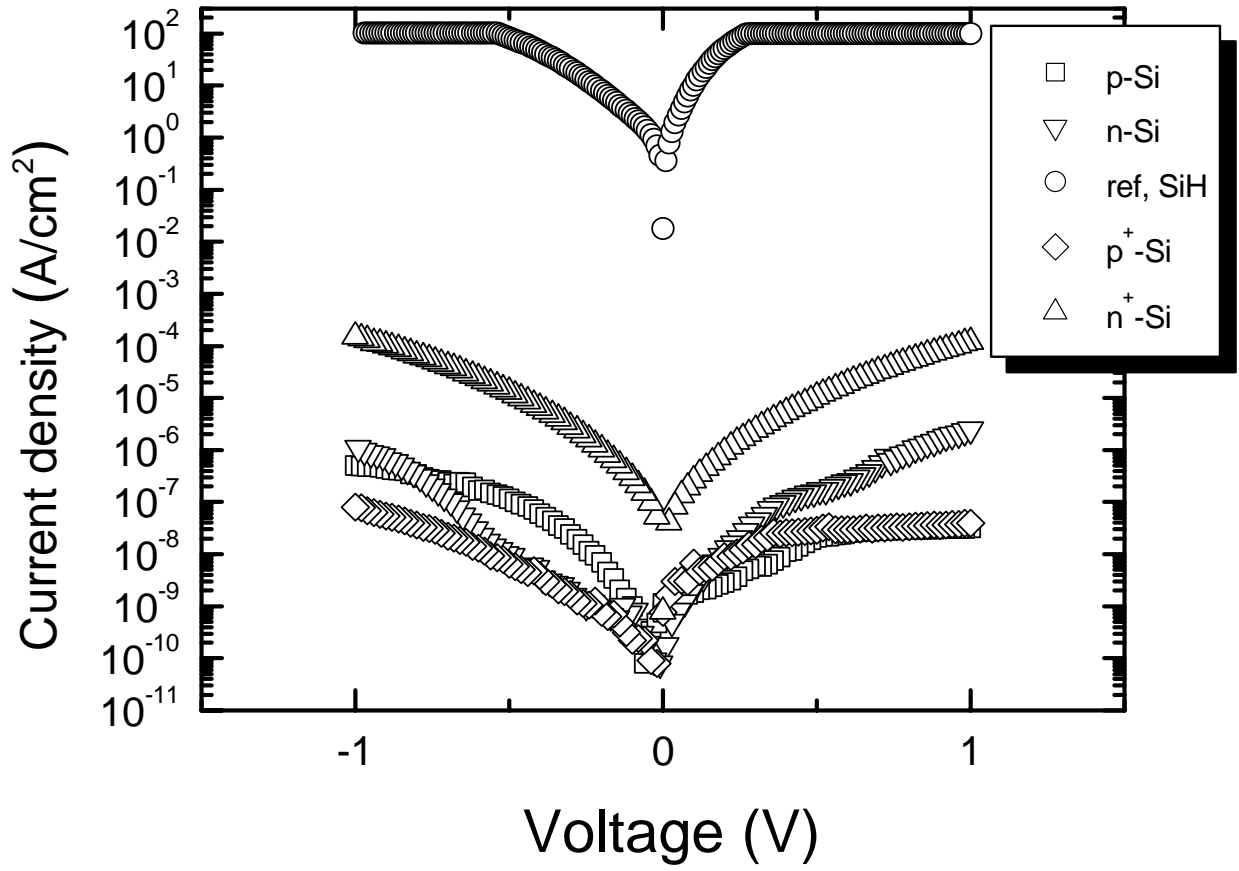

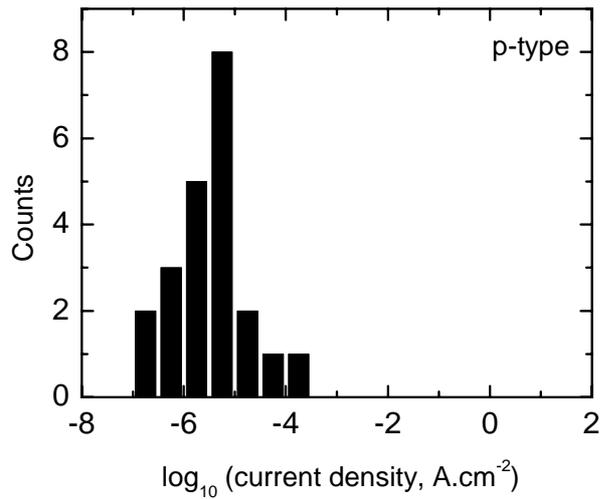

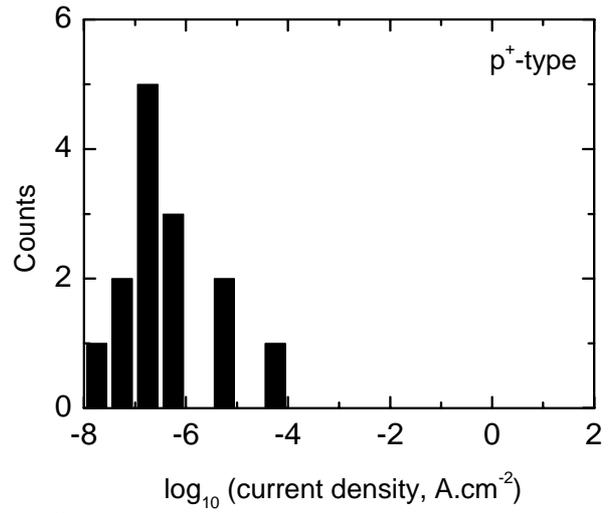

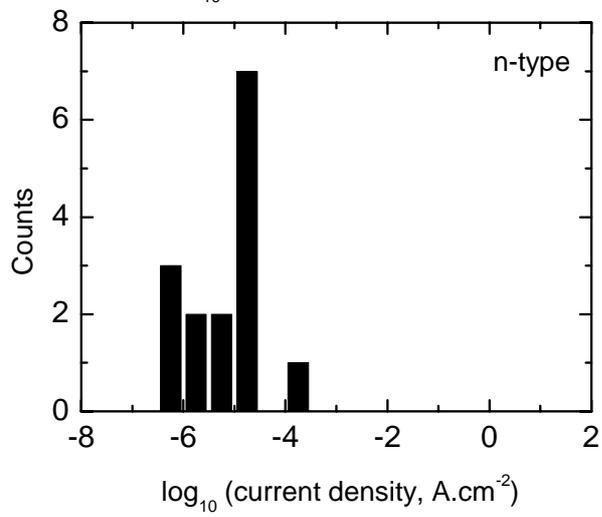

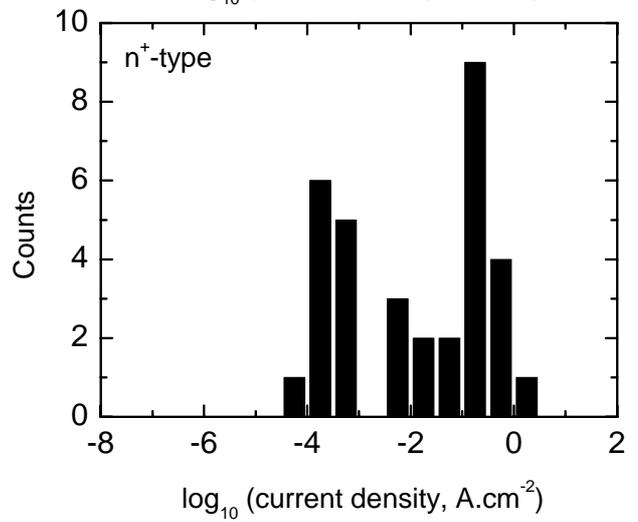

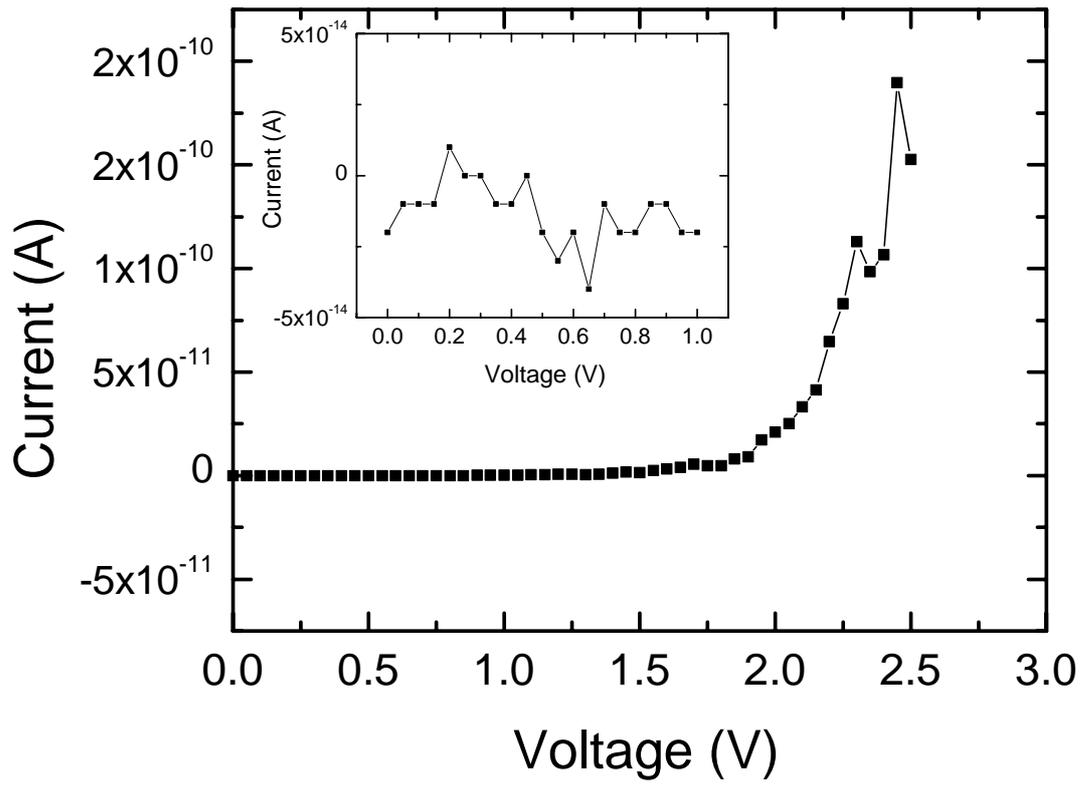

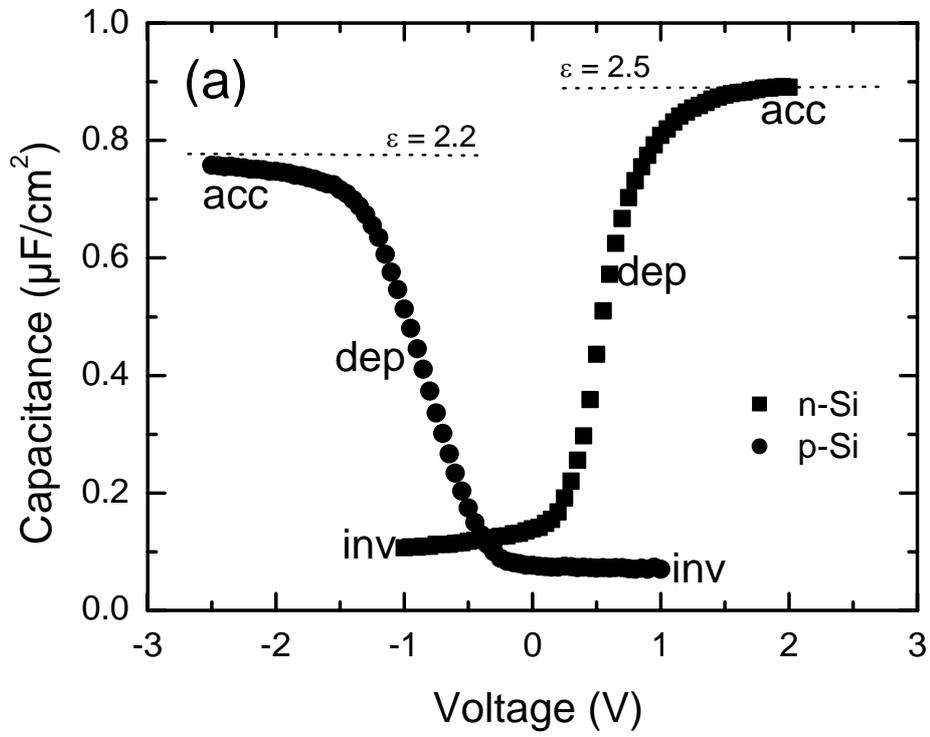

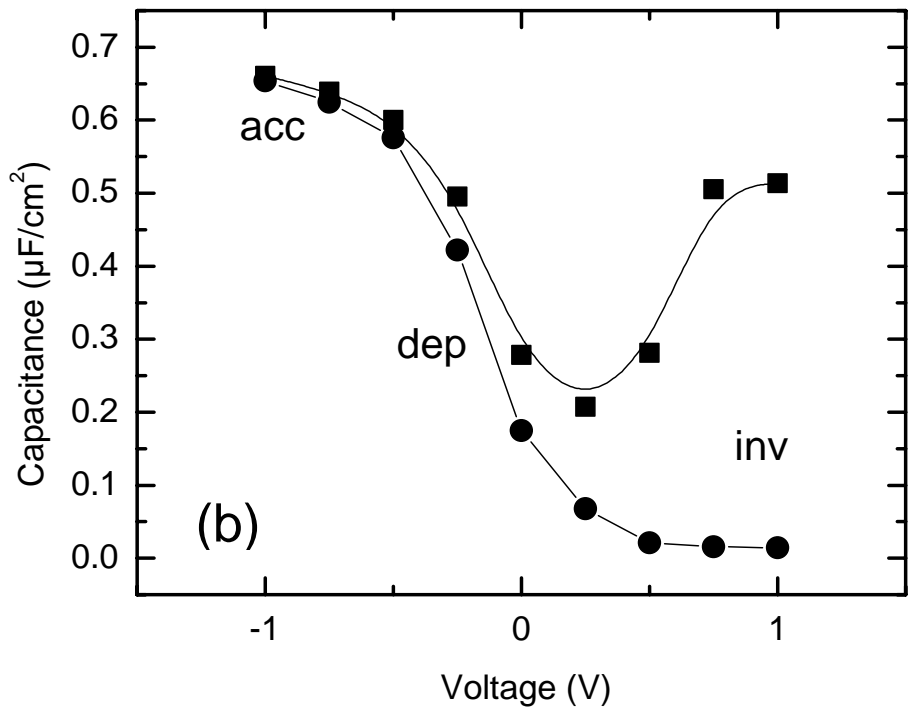